# Capacitive Sensing of Intercalated H$_2$O Molecules Using Graphene

Eric J. Olson,[†§] Rui Ma,[†§] Tao Sun,[‡] Mona A. Ebrish,[†] Nazila Haratipour,[†] Kyoungmin Min,[‡] Narayana R. Aluru,[‡] and Steven. J. Koester[†]*

[†]*Department of Electrical and Computer Engineering, University of Minnesota Twin Cities, 200 Union St. SE, Minneapolis, MN 55455*
[‡]*Department of Mechanical Science and Engineering, Beckman Institute for Advanced Science and Technology, University of Illinois at Urbana-Champaign, Urbana, Illinois 61802, USA*
*Ph: (612) 625-1316, FAX: (612) 625-4583, Email: skoester@umn.edu

## Abstract

Understanding the interactions of ambient molecules with graphene and adjacent dielectrics is of fundamental importance for a range of graphene-based devices, particularly sensors, where such interactions could influence the operation of the device. It is well-known that water can be trapped underneath graphene and its host substrate, however, the electrical effect of water beneath graphene and the dynamics of how it changes with different ambient conditions has not been quantified. Here, using a metal-oxide-graphene variable-capacitor (varactor) structure, we show that graphene can be used to capacitively sense the intercalation of water between graphene and HfO$_2$ and that this process is reversible on a fast time scale. Atomic force microscopy is used to confirm the intercalation and quantify the displacement of graphene as a function of humidity. Density functional theory simulations are used to quantify the displacement of graphene induced by intercalated water and also explain the observed Dirac point shifts as being due to the combined effect of water and oxygen on the carrier concentration in the graphene. Finally, molecular dynamics simulations indicate that a likely mechanism for the intercalation involves adsorption and lateral diffusion of water molecules beneath the graphene.

[§] Equal contribution

Keywords: graphene, sensor, varactor, capacitance, water

The successful exfoliation of single-layer graphene and subsequent development of chemical vapor deposition (CVD) for producing large-area graphene sheets has resulted in many interesting device applications. Its use in field-effect transistors,[1] mixers,[2] optical modulators,[3] photodetectors,[4] and a wide variety of chemical sensors is of particular note.[5-9] In nearly all of these device concepts, the intimate interactions between graphene and adjacent dielectrics is critical, yet has not been explored in detail. For instance, it has been shown previously, using density functional theory (DFT) simulations, that the equilibrium distance between graphene and $HfO_2$ is 0.30 nm.[10] However, it has also been shown that this equilibrium distance can change in the presence of defects on graphene or in the adjacent dielectrics, as the defects create bonding sites that result in stronger coupling between graphene and $HfO_2$.[10] The intimate surface interactions can become even more complex with the introduction of small molecules such as $H_2O$,[11] which can often be trapped between the graphene and the adjacent surface. Previously, atomic-force-microscopy (AFM) studies have shown that exfoliated graphene on mica can visualize the trapped water underneath due to the displacement of graphene.[12] However, to date, these trapped molecules have only been probed using physical analysis techniques such as AFM.[12,13] It would be extremely useful if such molecular interactions could be probed using electrical techniques, as such methods could allow a greatly improved understanding of the dynamics of these processes.

We have recently proposed a capacitance-based wireless sensor concept based upon a metal-oxide-graphene (MOG) variable capacitor (varactor),[14] and demonstrated its operation using water vapor as a test analyte.[15] These capacitors use a local buried gate electrode over which a thin layer of $HfO_2$ is deposited, followed by CVD-grown graphene on top.[16] The total capacitance, $C_{TOT}$ of such a device equals $(C_{OX}^{-1} + C_Q^{-1})^{-1}$, where $C_{OX}$ is the oxide capacitance and $C_Q$ is the quantum capacitance of graphene.[17] If $C_{OX}$ is sufficiently large, the capacitance varies vs. applied voltage and has a minimum when the Fermi level is at the Dirac point. Because these devices utilize a back-gated geometry, the surface of the graphene is exposed to the local environment and is therefore sensitive to changes of the chemical composition of the surroundings. In previous experiments, the resonant frequency of an LC circuit consisting of a graphene varactor coupled to an inductor was shown to reproducibly change in response to relative humidity (RH).[15] Though the physical nature of the interaction between water and the graphene surface was not immediately clear, it was speculated that the capacitance change was due to a Dirac point shift, as has been observed in resistive graphene sensors.[8,9] However, more recent experiments suggest that water intercalated between $HfO_2$ and graphene may also affect the capacitance behavior,[18] though these experiments were not performed under controlled atmospheric conditions.

In this work, we demonstrate unambiguously that water intercalates between graphene and $HfO_2$ and that this mechanism can be probed using a capacitance-based technique which shows the process is both reversible and fast. These capacitance measurements further demonstrate that increasing RH produces an n-type doping shift, contrary to typical explanations of a p-type shift normally associated with water. The $H_2O$ intercalation is confirmed using atomic force microscopy (AFM) measurements. The capacitance change associated with increasing RH is in agreement with the expected change in thickness and dielectric constant of the van der Waals gap between graphene and $HfO_2$ as determined using DFT simulations. In addition, the doping effects attributed to the combined effect of water and oxygen molecules are also supported by DFT results. Finally, molecular dynamics (MD) simulations confirm the feasibility of the proposed lateral diffusion mechanism for $H_2O$ intercalation. These results provide fundamental new insight into molecular interactions in the technologically important graphene/HfO$_2$ system and could open the door to a new sensing transduction mechanism.

**Device Fabrication and Capacitance Measurements**

The graphene-based capacitive sensing devices used here were similar to those described previously.[18] The full fabrication sequence is detailed in the Methods section and is briefly described as follows. The device structure consisted of a planarized buried metal gate electrode embedded in a thick thermal SiO$_2$ layer grown on a Si wafer. On top of the gate electrode, a HfO$_2$ layer with equivalent oxide thickness (EOT) of ~ 3.5 nm was deposited by atomic layer deposition (ALD) followed by transferring and patterning of single-layer CVD-grown graphene. Finally, metallic Ohmic contacts were made to the graphene. These devices were then mounted and wire-bonded to a header that was placed in a chamber for testing under a variety of atmospheric conditions. It is important to note that the device was configured in such a way that the entire graphene region above the gate electrode was exposed to the ambient atmosphere, as depicted schematically in Figure 1a. The RH inside the chamber was controlled by carefully adjusting the flow rates of water-saturated air and desiccated air into the chamber. As the atmosphere was being cycled between high and low humidity, the capacitance vs. voltage (C–V) characteristic of the MOG capacitor was continuously monitored with an Agilent B1500A semiconductor parameter analyzer, and roughly 500-1000 C–V measurements were performed in a typical humidity cycling experiment. It should be pointed out in this analysis that RH values above ~ 80% were excluded from our capacitance-based analysis in order to avoid parasitic effects associated with condensation on the sample surface and adjacent bond wires.

For the first series of measurements, the device under test consisted of an MOG capacitor with a single gate finger with length and width of 40 μm and 40 μm, respectively. Typical C–V characteristics of this device are shown in Figures 1b and 1c, and correspond to the points on the RH time sequence shown in Figure 1d. The plot in Figure 1b shows the characteristic at RH = 74% and 2.8% while the plot in Figure 1c shows the C–V curve when RH = 44% and 0.6%. There are three general trends observed in a high humidity environment compared to low humidity conditions. First, the maximum capacitance (measured at a gate voltage, $V_G = -3$ V) increases with higher humidity. Secondly, the average Dirac voltage between forward and reverse gate voltage sweeps, $V_{\text{Dirac-avg}}$, shifts toward more negative voltages, indicating graphene is less p-type doped when humidity is high compared to dry conditions. Concurrently, the hysteresis, defined as the Dirac voltage difference between forward and reverse gate voltage sweeps, $\Delta V_{\text{Dirac}}$, also increases with increasing humidity. Finally, the ratio between the maximum and minimum capacitance, $C_{\max}/C_{\min}$, becomes larger with increasing humidity.

Figures 1e-f show the time evolution of $C_{\max}$, $C_{\max}/C_{\min}$, for the forward voltage sweep, and Figures 1g-h show the time sequence of $V_{\text{Dirac-avg}}$ and $\Delta V_{\text{Dirac}}$. The plot in Figure 1e indicates that the $C_{\max}$ value tracks the RH closely, though a very slow drift toward higher capacitance values is observed, particularly for low RH values. In particular, the initial $C_{\max}$ under dry conditions is 7.24 pF at the start of the humidity cycling sequence, while at the end of the sequence (over 400 minutes later) the dry $C_{\max}$ has increased to 7.45 pF. This behavior is consistent with the drift in resonant frequency observed in our previously published wireless sensing experiments.[15] Surprisingly, the $C_{\max}/C_{\min}$ ratio increases with increasing humidity, and has remarkably large values (> 1.6), approaching the highest values reported in the literature to date.[18,19] Perhaps most surprisingly, the negative shift of $V_{\text{Dirac-avg}}$ in Figure 1g at high RH values, indicates that graphene is less p-type doped due to the effect of water molecules. This is contrary to the fact that graphene has been observed to shift more p-type in the presence of $H_2O$, which has been previously attributed to doping of graphene by water adsorbing onto residue or defect sites in graphene.[20,21] Finally, as shown in Figure 1h, the hysteresis increases with RH, roughly in proportion to humidity.

The results in Figure 1 suggest water intercalation between graphene and $HfO_2$ with increasing humidity as the source of the observed behavior. This hypothesis is supported by the fact that the increase in maximum capacitance can only be explained by an increase in the effective $C_{OX}$ with higher RH. Even though intercalation of water into the interfacial layer is expected to result in a larger gate-to-graphene separation, the larger dielectric constant of the interfacial water is expected to decrease the EOT when compared to the case where a vacuum gap exists. The increase in the hysteresis with increasing humidity is also consistent with trapped moisture underneath the graphene.[22,23] Finally, the observation that the maximum capacitance does not return to its original value after the full humidity cycling sequence is consistent with trapped moisture beneath the graphene.

**Atomic Force Microscopy Analysis**

To explore the hypothesis of water infiltration into the interfacial layer, pulsed force mode AFM was utilized to measure the step height between the $HfO_2$ dielectric and the graphene as a function of RH. Direct verification of water infiltration into the interfacial layer between mica and graphene has previously been performed by AFM;[12,24,25] however those studies utilized exfoliated graphene which avoids structural defects and residue resulting from transfer of CVD-grown graphene. To verify that water was indeed intercalating into the space between graphene and $HfO_2$, in our experiment, AFM imaging was performed on a sample of CVD-grown graphene that had been transferred onto a Si substrate containing 300 nm thermally-grown $SiO_2$ followed by ALD-deposited $HfO_2$. Because we have previously observed that PMMA residue congregates at graphene defect sites, the graphene was first imaged by contact mode under rather high tip force, since this process mechanically creates small holes in the graphene that are free of PMMA residue. It is also typical in the fabricated varactors that small tears occur due to non-idealities in the transfer process and growth which provide "access points" for water intercalation.[18] Since it is difficult to locate these defects during the AFM measurement, the intentional holes in the graphene provide a reasonable method to reproduce these naturally occurring defects. To clearly show the intercalation of water into the interfacial layer, the hole shown in Figure 2a was repeatedly imaged using pulsed force mode AFM[26] at varied relative humidity. Pulsed force mode was utilized for imaging to avoid further tearing of the graphene by imaging with contact mode and the introduction of imaging artifacts created by attractive-repulsive regime switching that has previously been observed when imaging graphene-oxide boundaries by tapping mode AFM.[27]

After imaging, a histogram of the heights contained within the area highlighted in Figure 2a was extracted. An example histogram taken in high humidity conditions is shown in Figure 2b, while a histogram taken in a low humidity environment is shown in Figure 2c. The two peaks were then fit to two Gaussian distributions corresponding to the substrate and graphene heights and the step height from the oxide substrate to graphene was calculated as the distance between the two peaks. This procedure was needed due to the inherent roughness of both the underlying $HfO_2$ as well as the residual PMMA reside on the graphene. While much sharper images have previously been produced from AFM studies of graphene on mica,[12] for this analysis, it was important to perform the studies using a structure representative of the realistic varactors geometry. The extracted step height, as well as the chamber humidity, with respect to the sequence in which the images were acquired is plotted in Figure 2d. These images allow us to conclude that water is intercalating into the interfacial layer between graphene and $HfO_2$ and that the intercalation gives rise to an increased separation. However, since the dielectric constant of the water is expected to be considerably higher than vacuum,[28] the capacitance should increase with increasing interfacial moisture. Indeed, Figure 2e shows that the change in the EOT (extracted by fitting the data in Figure 1 using the procedure outlined in reference 18) shows good correlation with the AFM data, with the EOT decreasing by roughly 2.5 Å for every 1 Å increase in step height. Further quantitative correlation between the EOT and graphene step height changes, are described later in this manuscript. This data further supports the water infiltration as the likely mechanism for higher capacitance with increasing humidity. A slow drift of the base step height in the AFM analysis was also observed when cycling the relative humidity, as shown in Figure 2d. This result supports the hypothesis described previously that residual trapped moisture is the primary cause of the capacitance drift observed in Figure 1f. Further studies are needed to determine whether or not graphene-on-$HfO_2$ devices can be made fully reversible.

While water intercalation into the interfacial layer directly explains the increase in capacitance observed at high humidity, the source of the increase in $C_{max}/C_{min}$ in the MOG devices at high humidity compared to low humidity is less clear. We have previously analyzed changes in the $C_{max}/C_{min}$ ratio in MOG structures[18] and attributed this change to a decrease in the disorder associated with random potential fluctuations. One possible source of the disorder may be learned from the AFM results. We have observed that the $HfO_2$ layer in the device exhibits a substantially higher roughness than the graphene does, which suggests that graphene is bridging gaps over peaks in the oxide. Upon introduction of water to the interfacial region, the hygroscopic nature of the $HfO_2$ draws water into the interface. If a homogenous layer of water physically fills the gaps between graphene and $HfO_2$, the variability of the environment immediately below the graphene is thus reduced.

**Effect of Ambient Oxygen**

While the increase in $C_{max}$ and $C_{max}/C_{min}$ with humidity can reasonably be explained by intercalation of water beneath the graphene, the dependence of $V_{Dirac-avg}$ on humidity is much less obvious. Therefore, in order to develop additional insight into this problem we performed humidity sensing experiments where the carrier gas was replaced by dry nitrogen. This experiment was performed to help understand the effect of oxygen on the device response. These experiments were performed on a multi-finger (8 × 5 μm × 40 μm) MOG varactor and the results are shown in Figure 3. Here, two separate RH sweeps were performed where in one case (Figure 3a), the carrier gas consisted of desiccated air and in the other (Figure 3f), dry $N_2$ was the carrier gas. Then, $C_{max}$, $C_{max}/C_{min}$, $V_{Dirac-avg}$, and $\Delta V_{Dirac}$ (for both forward and reverse voltage sweeps) were monitored as a function of time. While the general trends for $C_{max}$ and $C_{max}/C_{min}$ were found to be relatively similar, $V_{Dirac-avg}$ value was found to be substantially more positive for air than nitrogen. It should be pointed out that a relatively large drift in $V_{Dirac-avg}$ was observed under $N_2$ ambient conditions, which could be due to the slow desorption of oxygen molecules from the $HfO_2$-graphene interface. Even considering this drift, the values of $V_{Dirac-avg}$ for air were still about 0.3-0.5 V larger than those for nitrogen at the same humidity. This result implies that the $O_2$ in air has a p-type doping effect on graphene. Plots of capacitance vs. voltage corresponding to the data points on the RH vs. time plot in Figures 3a and 3f are shown in Figure S4 of the Supplementary Information. In this measurement, the nitrogen data was taken after the air experiment, but very similar results were observed when the order of the experiments was reversed, which is shown in Figure S5 of the Supplementary Information.

**Density Functional Theory and Molecular Dynamics Simulations**

To better understand the above results, we performed DFT and MD simulations for the interactions most relevant to our devices. Details of the simulation setup are described in the Methods section.

*System configurations during simulations*

Figure 4 shows the atomic structure and corresponding charge redistributions of the graphene-$HfO_2$ system for different steps of the experiment. Figure 4a represents the system of graphene above pristine $HfO_2$, which serves as a reference to clearly show the doping effect of graphene in the following processes. To mimic the vacuum condition in the experiments, in Figure 4b, graphene was placed above $HfO_2$ with 4 oxygen vacancies (VO4). It is clear that the charge redistribution area between graphene and $HfO_2$ is larger than the pristine system in Figure 4a, and graphene has a larger area of electron accumulation due to $HfO_2$ defects, indicating n-type doping. This is consistent with previous studies.[10] Figure 4c represents the structure in which 4 $O_2$ molecules are intercalated underneath graphene while 5 $O_2$ molecules are placed above graphene, which mimics the dry air condition in the experiments. It should be noted that $O_2$ underneath graphene fills the VOs by forming Hf-O bonds, which can eliminate the n-type doping effect originated from defective $HfO_2$. At the same time, $O_2$ molecules above graphene act as electron acceptors.[29,30] As a result, the charge redistribution area between graphene and $HfO_2$ in Figure 4c is much smaller than that in Figure 4b, while the charge redistribution area between graphene and $O_2$ is very large. The large area of electron depletion on graphene surface indicates electrons are transferred to $O_2$, hence graphene becomes p-type doped. Finally, to mimic the experimental situation where the device is exposed to humidity, in Figure 4d, 13 $H_2O$ molecules are placed between graphene and the $O_2$-filled $HfO_2$; in addition, 2 of the 5 $O_2$ above graphene are replaced with 2 H$_2$O, compared with the system in Figure 4c. From the corresponding charge redistribution, the electron depletion area on the graphene surface is smaller than in the case of Figure 4c, implying that fewer electrons are transferred from graphene to O$_2$, thus graphene becomes less p-type doped.

*Partial density of states and Bader population analysis*

For a more detailed understanding of the results presented above, the partial density of states (PDOS) of graphene for all the systems in Figure 4 was calculated and the results are shown in Figure 5a. The location of the minima of the PDOS of graphene represents the value of $E_{\text{Dirac}}$-$E_{\text{Fermi}}$ of graphene, reflecting the relative position of the Dirac Point and the Fermi Level. For graphene in Figure 4a, the minima of the PDOS locates close to 0 eV, indicating that the Fermi Level is almost equal to the Dirac Point and graphene is intrinsic; for graphene in Figure 4b, the minima locates in the negative area, which implies that the Fermi Level is above the Dirac point and graphene is n-type doped; similarly for Figure 4c, the minima is at positive values, implying that the Fermi Level is below the Dirac Point and graphene is p-type doped. Finally, for graphene in Figure 4d, the minima still locates in positive region, but moves to the left with respect to the minima for the situation in Figure 4c, implying that graphene is less p-type doped. The $E_{\text{Dirac}}$-$E_{\text{Fermi}}$ values corresponding to the minimum PDOS are summarized in Table 1.

For further evidence of the doping effect of graphene, Bader populations[31-34] of graphene and the number of electrons transferred to graphene for all the systems in Figure 4 were calculated and summarized in Table 1. The results show that graphene gains electrons (4.0929*e*) and so is n-type doped for the situation in Figure 4b with oxygen vacancies in HfO$_2$, loses electrons (−0.6004*e*) and so is p-type doped with the introduction of O$_2$ corresponding to Figure 4c, and loses fewer electrons (−0.3357*e*) and so is less p-type doped in the presence of both O$_2$ and H$_2$O, as in Figure 4d. In addition, Table 1 also lists the calculated interlayer distance between graphene and $HfO_2$. The results show that oxygen vacancies decrease the interlayer distance (1.72 Å) compared to the case of pristine from $HfO_2$ (2.01 Å), indicating a stronger interaction between graphene and defective $HfO_2$. When $O_2$ fills the VOs, the interlayer distance is increased to 1.88 Å, indicating a weakened interaction compared to the situation without $O_2$. Finally, when $H_2O$ intercalates as shown in Figure 4d, the interlayer distance increases substantially to 2.44 Å.

*Molecular dynamics simulations*

A final outstanding question in the experimental results is the process by which intercalation of $H_2O$ beneath graphene can occur, particularly under high humidity. MD simulations were utilized to understand this behavior. Figure 5b is a snapshot from the MD simulation at the point when water molecules are trapped between graphene and $HfO_2$. Water molecules were initially placed above graphene, and as the simulation evolves, these molecules migrate in between graphene and $HfO_2$. Due to water intercalation, the step height between graphene and $HfO_2$ increased by about 1.18 Å, which is comparable to the experiment result determined by AFM.

**Summary**

In summary, we have shown that water intercalation into the interfacial layer between graphene and $HfO_2$ is reversible with repeated humidity cycling and can be detected electrically using a metal-oxide-graphene device structure. AFM results confirm the intercalation process and agree quantitatively with the capacitance-voltage characteristics. The mechanism is further explained using density functional theory, which also provides a description for the observed doping effect as being due to the combined interaction of $H_2O$ and $O_2$. This work has wide-ranging implications for the fundamental understanding of the effect of $H_2O$ on sensors (both as a target molecule and an interfering specie), as well as a wide range of electronic devices based upon graphene.

**METHODS**

*Fabrication of metal-oxide-graphene devices*

Graphene varactors were constructed by following the procedure outlined in reference 16. The fabrication started by growing a thick layer of $SiO_2$ (980 nm) on Si by thermal oxidation. The gate electrode patterns were then and exposed and developed using photolithography, and the $SiO_2$ etched with buffered oxide etch, and Ti/Pd (10/40 nm) deposited by evaporation and lifted off. A thin layer (9.8 nm) of $HfO_2$ was then deposited by atomic-layer deposition (ALD) at 300 °C and annealed at 400 °C for 5 minutes. Single-layer graphene grown by chemical vapor deposition (CVD) was then transferred onto the wafer using the procedure described below and patterned by photolithography and etched with $O_2$ plasma. Lastly, Ti/Pd/Au (1.5/35/70 nm) metallic contacts to the graphene were patterned and lifted off. The test chips contained a variety of single- and multiple-gate-finger varactors such as those described in reference 16. The fabricated devices were then affixed and wire-bonded to either a 32-pin or a 40-pin header.

The steps in the graphene transfer process were as follows. First, CVD was used to grow single-layer graphene onto a copper foil. Then PMMA (4% in chlorobenzene) was spin-coated onto the foils at 3000 rpm for 60 sec and then the sample was baked at 180 ºC for 10 minutes on a hot plate. Next, the graphene on the backside of the foil was removed by etching in an $O_2$ plasma for 15 seconds. The sample was then soaked in a bath of ammonium persulfate overnight to remove the copper and then rinsed twice in DI water for 10 minutes. Then the graphene was transferred onto the substrate with the pre-patterned gate electrodes and $HfO_2$ dielectric using an aqueous transfer method. Finally, the sample was baked at 65 ºC for 10 minutes and 180 ºC for 15 minutes, and finally the PMMA was removed using acetone and an IPA rinse.

*Measurement of humidity response*

Unless otherwise stated, all devices were placed in vacuum (~$10^{-6}$ Torr) for a minimum of 24 hours to desorb water from the device before measurement. The devices were then removed from the vacuum and immediately placed in a zero-insertion force socket inside a chamber similar to that used in reference 15, though with smaller overall volume and with BNC feedthroughs added to allow for wired measurements. Relative humidity inside the chamber was controlled by adjusting the relative flow rates of water-saturated and dry air. Water-saturated air was produced by bubbling compressed air through warm deionized water and dry air was produced by passing compressed air through calcium sulfate desiccant. An Electro-Tech Systems Model 514 humidity controller was used to measure the relative humidity in the chamber. Capacitance vs. voltage (C–V) characteristics were then measured with an Agilent B1500A semiconductor parameter analyzer.

*Atomic force microscopy*

A more detailed description of the capabilities and operational concerns for pulsed force mode atomic force microscopy (PFM AFM) is available in reference 35. Here, PFM AFM was performed at ambient temperature using an Agilent 5550 environmental scanning probe microscope equipped with Witec digital pulsed force mode electronics and a MikroMasch NSC-36 cantilever with an 8 nm tip radius and nominal spring constant of 1 N/m. For all measurements, an oscillation frequency of 2000 Hz while recording one line per second and 512 pixels per line. As a result, each pixel is the result of two approach–retract cycles. The loss angle and adhesion force for each pixel were monitored in addition to the overall topography of the sample. The humidity in the AFM chamber was controlled by an Electro-Tech Systems Model 514 humidity controller in conjunction with an Electro-Tech Systems humidifier and a mechanical air pump to deliver humidified and desiccated air, respectively.

*Density Functional Theory and Molecular Dynamics simulations*

In order to investigate the electronic property change during the operation of the MOG devices in the presence of humidity, systematic first principle density functional theory (DFT) simulations were performed using the SIESTA package.[36] We employed generalized gradient approximations (GGA) with Perdew-Burke-Ernzerhof (PBE) exchange-correlation functional.[37] Double zeta polarized basis set (DZP) was used. The real space grid mesh cutoff value was 300 Ry. A Monkhorst-Pack grid of $4 \times 4 \times 1$ was used for structure relaxation and $18 \times 18 \times 1$ for electronic property calculations.[38] A vacuum region of about 15 Å was applied in the z-direction of the periodic box to avoid nonphysical effect from the images. The structures were relaxed until the maximum residual force of the system was less than 0.05 eV/Å before obtaining the electronic properties.

To understand water intercalation behavior between graphene and $HfO_2$, molecular dynamics (MD) simulations were performed using the LAMMPS package.[39] We applied a hybrid potential to describe atomic interactions. The graphitic carbon-water non-bonded parameters[40] were used for the interaction between graphene and water molecules. The carbon atomic interaction within graphene was modeled by AIREBO potential.[41] During the simulation, the positions of $HfO_2$ substrate were fixed, and the van der Waals force [described by Leonard Jones (LJ) potential] and Coulomb force (described by Coulomb potential) were used to model the interactions between $HfO_2$ and graphene/water. LJ parameters for Hf-Hf and O-O interactions were obtained from Universal Force Field (UFF),[42] and then a mixing rule was used to obtain LJ parameters modeling the interactions of Hf-C, Hf-O and Hf-H. The time step was set to be 0.1 fs, and NVT ensemble was applied with Nosé-Hoover thermostat at 300 K. We applied periodic boundary condition (PBC) in x- and y-directions for $HfO_2$, while only in y-direction for graphene. The size of graphene in x-direction was smaller than that of $HfO_2$ substrate, so that water molecules could intercalate underneath graphene.

Amorphous $HfO_2$ (a-$HfO_2$) was used as a substrate to support graphene. We generated a-$HfO_2$ structure from MD simulation. Charge optimized many body (COMB) potential[43] was used to illustrate the interaction between hafnium (Hf) and oxygen (O) atoms. The time step was set to be 0.1 fs, and NPT ensemble was applied with Nosé-hoover thermostat at 300 K and barostat at 1 atm. We applied typical annealing and quenching processes to generate amorphous structure of $HfO_2$. $HfO_2$ monoclinic structure was first constructed and relaxed at 300 K, and then the system temperature was gradually increased to 4000 K and maintained. After annealing, the system was gradually cooled down to 300 K, and maintained. The structure of α-$HfO_2$ was relaxed again in DFT before calculating the electronic properties.

To obtain the change in electron density of graphene due to its interaction with surrounding substances, the charge density redistribution ($\Delta\rho$) was calculated. It is defined as

$$\Delta\rho = \rho_{total} - \Sigma\rho_{sub}, \tag{1}$$

where $\rho_{total}$ and $\rho_{sub}$ represent the charge density distribution of the whole system and all isolated subparts, respectively. For example, for the system of graphene with water molecules on top and $HfO_2$ substrate underneath, the charge density redistribution is

$$\Delta\rho = \rho_{total} - \rho_{graphene} - \rho_{H_2O} - \rho_{HfO_2}. \tag{2}$$

After obtaining the results, the charge density redistribution was visualized using the XCrySDen program [44]. It is possible that during experiments in dry air, oxygen molecules can enter underneath graphene and be adsorbed on $HfO_2$ surface. To understand the nature of this adsorption, we calculated the binding energies between each $O_2$ and $HfO_2$ substrate. The binding energy ($E_{binding}$) was calculated as

$$E_{binding} = E_{total} - E_{HfO_2 + other\ O_2} - E_{specified\ O_2}, \qquad (3)$$

where $E_{total}$ is the energy of the total system, $E_{specified\ O_2}$ is the energy of one specified $O_2$ molecule and $E_{HfO_2 + other\ O_2}$ is the energy of $HfO_2$ substrate and other $O_2$ molecules.

**ACKNOWLEDGEMENTS**

This work was funded by the Minnesota Partnership for Biotechnology and Medical Genomics Decade of Discovery in Diabetes Program. Portions of this work were carried out in the College of Science and Engineering Characterization Facility, University of Minnesota, which has received capital equipment funding from the NSF through the UMN MRSEC program under Award Number DMR-1420013.

**AUTHOR CONTRIBUTIONS**

E.J.O., R.M., M.A.E. and S.J.K. designed and directed this study and analyzed the results. N.H. and M.A.E. performed device fabrication. E.J.O. and R.M. performed the sensor measurements. T.S. and K.M. performed computational simulations under the supervision of N.R.A. E.J.O., R. M., T.S. and S.J.K. wrote the manuscript.

**SUPPORTING INFORMATION AVAILABLE**

Supplementary Information is available in the online version of the paper. Correspondence and requests for materials should be addressed to S.J.K.

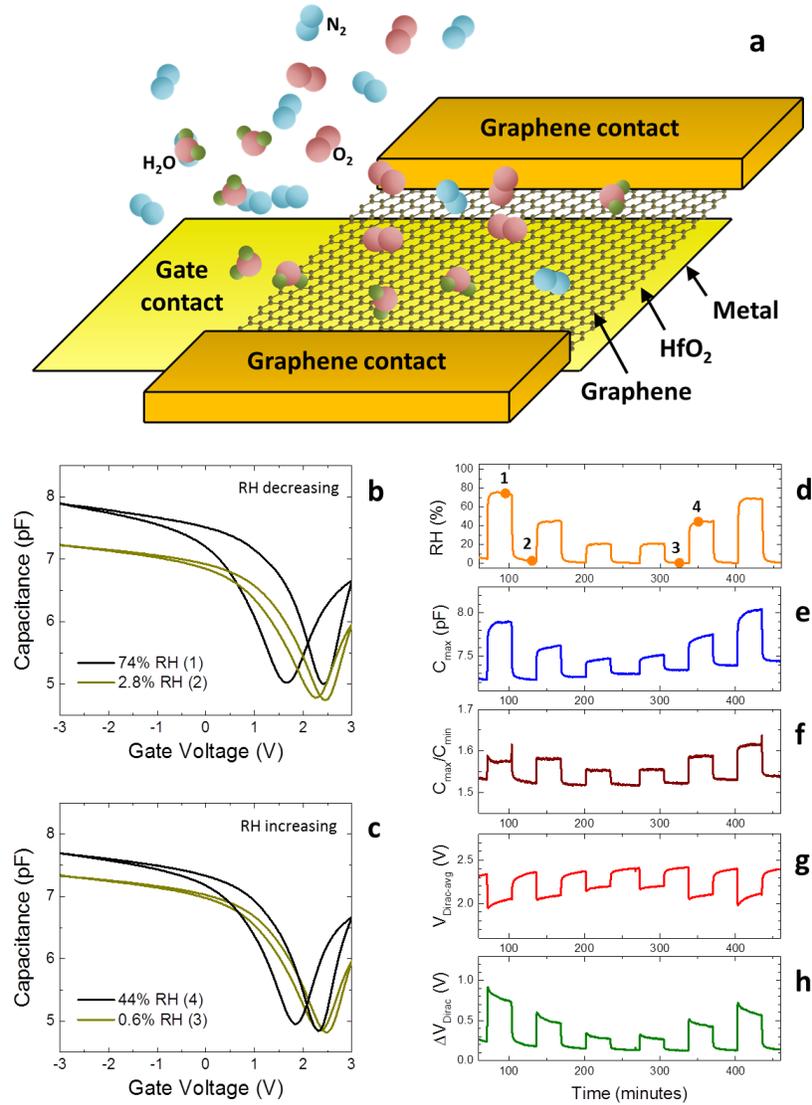

Figure 1. Diagram and sensing behavior of a metal-oxide-graphene (MOG) varactor at different values of relative humidity (RH). (a) Cartoon showing basic structure of the varactor and depicting primary ambient gas-phase molecules. The gate length and width are 40 μm and 40 μm, respectively, while the $HfO_2$ thickness is 9.8 nm. (b)-(c) Plot of capacitance vs. voltage for a graphene varactor for (b) decreasing and (c) increasing RH, where the plots correspond to the data points on the RH vs. time plot in (d). (e) Plot of maximum capacitance, $C_{max}$ for forward gate voltage sweep, (f) maximum to minimum capacitance ratio, $C_{max}/C_{min}$ for forward gate voltage sweep. (g)-(h) Average Dirac voltage, $V_{Dirac-avg}$, and the hysteresis between increasing and decreasing gate voltage sweeps, $\Delta V_{Dirac}$, vs. time corresponding to the RH sequence in (d).

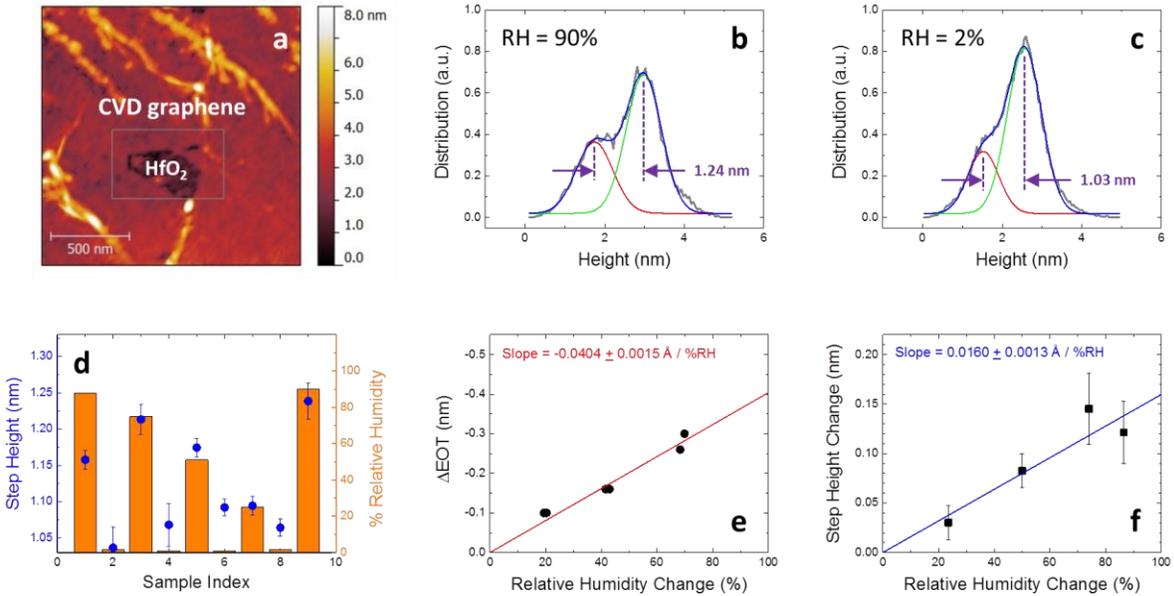

Figure 2. Data showing atomic force microscopy (AFM) measurements of CVD-graphene on $HfO_2$ and extracted step heights for different relative humidity (RH) levels. (a) AFM scan of CVD-grown graphene on $HfO_2$, where graphene has been removed using the AFM in a selected region of the sample. (b) Height distribution map for the indicated rectangular region in (a) for RH = 90%. The scans have been fit using two Gaussian distributions and where the difference in the peak heights is indicative of the step height between the graphene and $HfO_2$. (c) Height distribution map for the indicated rectangular region in (a) for RH = 2%. (d) Height of graphene flake relative to the $HfO_2$ for a sequence of measurements under different RH conditions. The step height is observed to increase with RH, though the step height also incurs a slow drift with repeated measurements. (e) Change in equivalent oxide thickness, ΔEOT, with ΔRH extracted by theoretical fitting of the data in Figures 1d-e. The points are the experimental data and the red line is the best fit straight line through the origin. (f) Plot of change in step height, Δh vs. change in relative humidity, ΔRH, where the step height change is based upon the difference between the AFM extracted step heights between successive measurements. The points indicate the experimental data and the blue line is the best fit straight line through the origin.

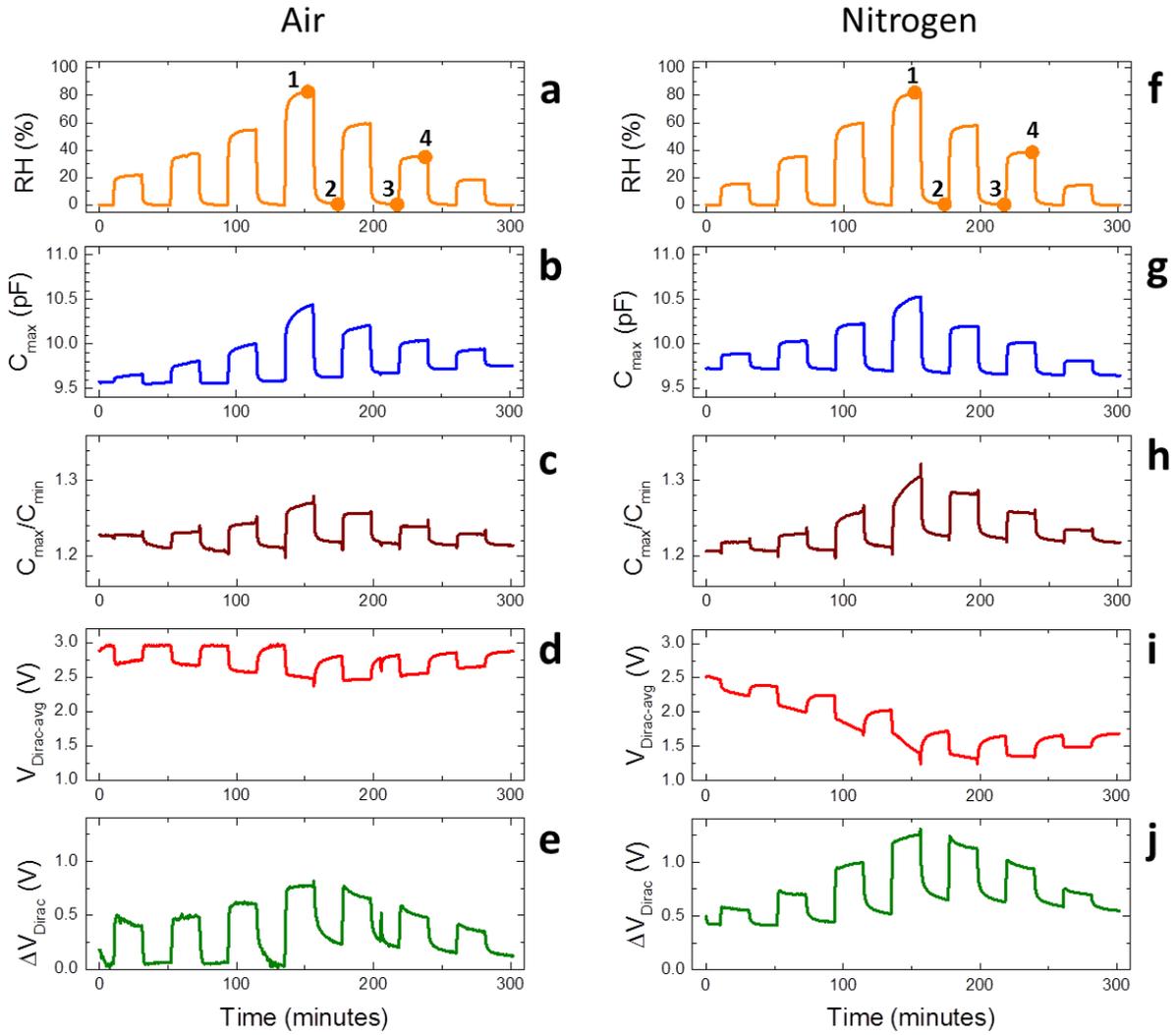

Figure 3. Comparison of MOG humidity sensing characteristics with air and nitrogen as the carrier gas (for a different device than the one in Figure 1). (a) Plot of relative humidity, (b) maximum capacitance, $C_{max}$ for forward gate voltage sweep, (c) maximum to minimum capacitance ratio, $C_{max}/C_{min}$ for forward gate voltage sweep, (d) Average Dirac voltage, $V_{Dirac-avg}$ between forward and reverse gate voltage sweeps vs. time and (e) The hysteresis, $\Delta V_{Dirac}$, between forward and reverse gate voltage sweeps vs. time with desiccated air as the carrier gas. (f)-(j) Plot of same parameters as in (a)-(e) for same device measured using dry $N_2$ as the carrier gas. The measurements in (a)-(e) were performed first, then $N_2$ was flowed through the chamber for 30 minutes before performing the measurements shown in (f)-(j).

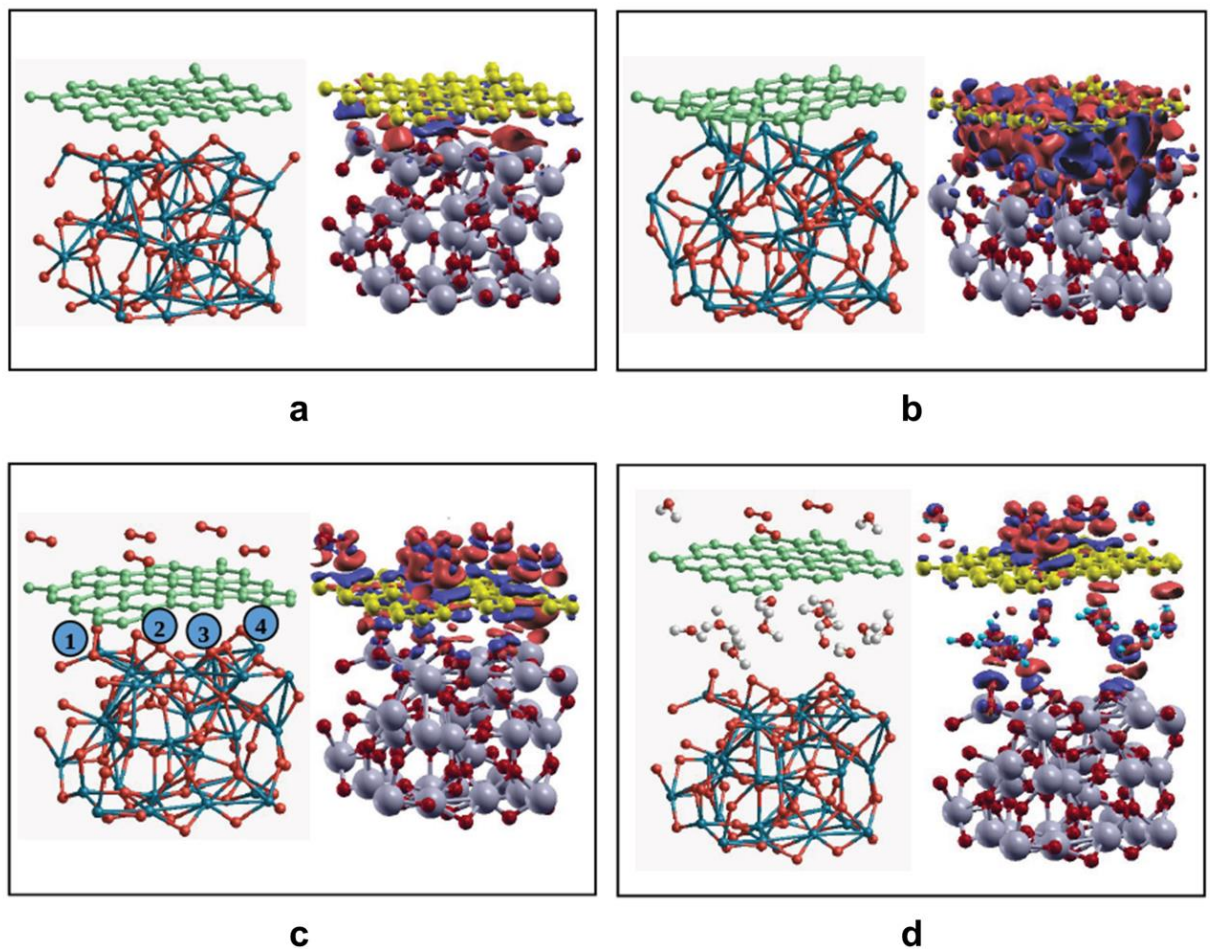

Figure 4. Atomic configurations (left) and charge redistributions (right) representing the range of conditions utilized in the sensing experiments. These configurations correspond to (a) graphene above pristine $HfO_2$, (b) graphene above $HfO_2$ with 4 oxygen vacancies (VO4), (c) graphene above $HfO_2$ (VO4) with $O_2(4)$ molecules in between (index 1, 2, 3 and 4) and $O_2(5)$ molecules on top, and (d) graphene above $HfO_2$ (VO4) with $O_2(4) + H_2O(13)$ molecules in between and $O_2(3) + H_2O(2)$ molecules on top. In atomic structures, green, red, blue, and white represent carbon, oxygen, hafnium and hydrogen, respectively. In charge redistribution plot, yellow, red, grey, cyan represent carbon, oxygen, hafnium, and hydrogen, respectively. The red and blue isosurfaces represent electron accumulation and depletion region, respectively. The isosurface value is set to be 0.001 $Å^{-3}$.

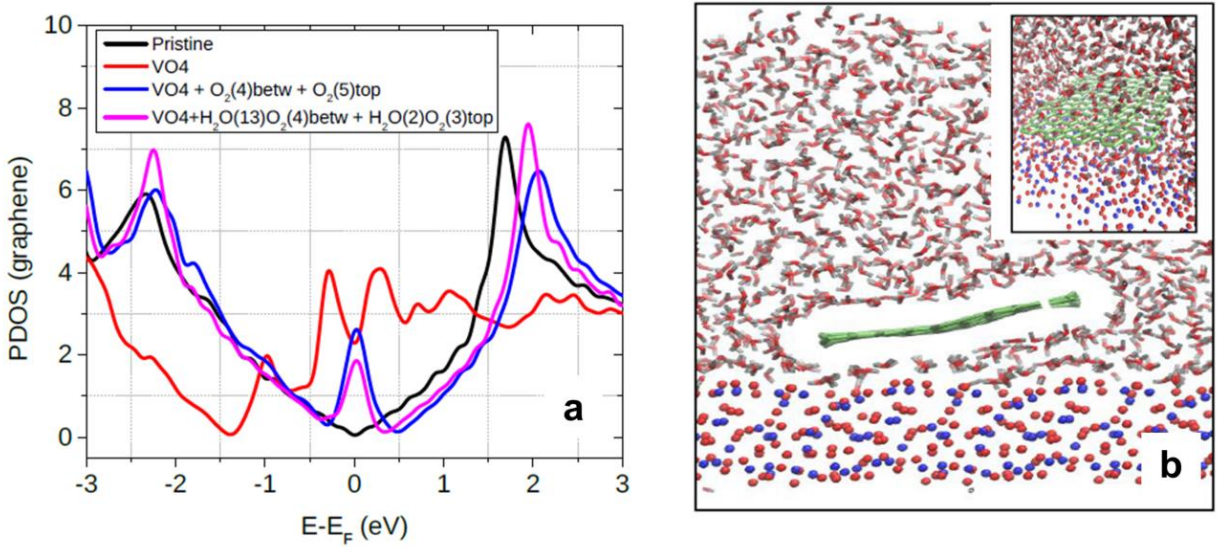

Figure 5. Plots showing results from density functional theory (DFT) and molecular dynamics simulations. (a) Partial density of states (PDOS) of graphene for: black: graphene above pristine $HfO_2$, red: graphene above $HfO_2$ with 4 oxygen vacancies (VO4), blue: graphene above $HfO_2$ (VO4) with $O_2(4)$ molecules between graphene and $HfO_2$ and $O_2(5)$ molecules on top of graphene, and magenta: graphene above $HfO_2$ (VO4) with $O_2(4)$ + $H_2O(13)$ molecules between graphene and $HfO_2$ and $O_2(3)$ + $H_2O(2)$ molecules on top. (b) Snapshot from MD simulations to describe water intercalation between graphene and $HfO_2$. The subplot is a 3D perspective view of the whole system. Green, red, blue, and grey represent carbon, oxygen, hafnium and hydrogen, respectively.

| Structure | Figure 4a | Figure 4b | Figure 4c | Figure 4d |
|---|---|---|---|---|
| Location of the minima of PDOS of graphene ($E_{Dirac}-E_{Fermi}$) (eV) | 0.00279 | -1.38409 | 0.48575 | 0.34018 |
| Bader population ($e$) | 191.9625 | 196.0929 | 191.3996 | 191.6643 |
| Charge transfer ($e$) | -0.0375 | 4.0929 | -0.6004 | -0.3357 |
| Doping effect | neutral | n-type | p-type | less p-type |
| Interlayer distance (Å) | 2.0107 | 1.7242 | 1.8788 | 2.4395 |

Table 1. Location of the minima of the PDOS of graphene, $E_{Dirac}-E_{Fermi}$, Bader population of graphene, amount of charge transferred to graphene, doping effect on graphene, and interlayer distance between graphene and $HfO_2$ for structures in Figure 4a to 4d.

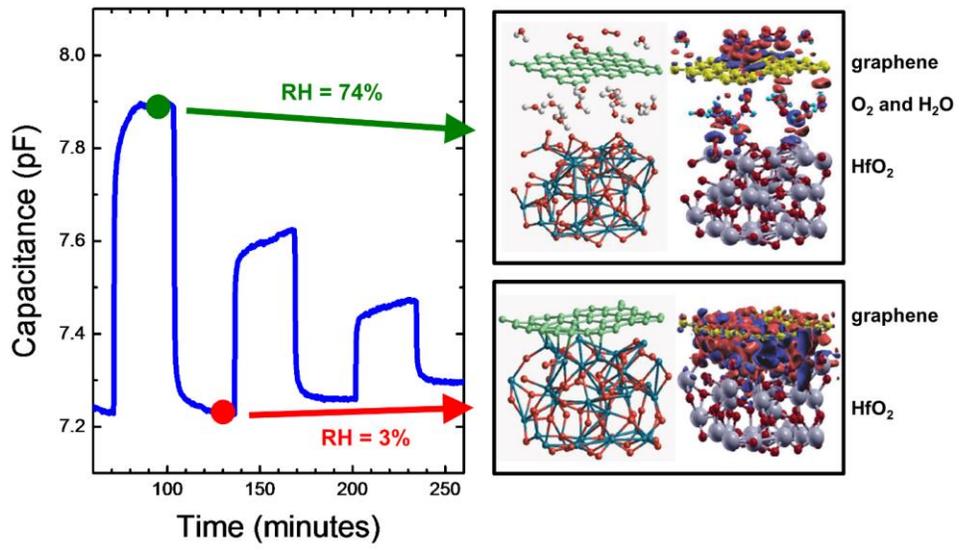

Table of Contents Figure